# A Primer on HIBS – High Altitude Platform Stations as IMT Base Stations

Sebastian Euler*, Xingqin Lin, Erika Tejedor*, Evanny Obregon*

*Ericsson

*Mobile communication via high-altitude platforms operating in the stratosphere is an idea that has been on the table for decades. In the past few years, however, with recent advances in technology and parallel progress in standardization and regulatory bodies like 3GPP and ITU, these ideas have gained considerable momentum. In this article, we present a comprehensive overview of HIBS – High Altitude Platform Stations as IMT Base Stations. We lay out possible use cases and summarize the current status of the development, from a technological point of view as well as from standardization in 3GPP, and regarding spectrum aspects. We then present some system level simulation results to shed light on the performance of HIBS. We conclude with pointing out several directions for future research.*

## Introduction

Over the past decades, mobile operators have greatly expanded the coverage of broadband wireless service, with the total number of mobile subscriptions exceeding 8 billion in 2021 [1]. Despite the wide deployment of terrestrial mobile networks, there is still a need for greater broadband connectivity services in remote communities. If non-terrestrial technologies could be deployed at a competitive cost, and if interworking with terrestrial networks could be achieved, they might have the potential of providing connectivity in remote areas, thereby complementing the terrestrial networks.

The term non-terrestrial networks (NTN) refers to networks utilizing spaceborne or airborne payloads for communication. The recent interest in spaceborne satellite communication has been centered on Low Earth Orbit (LEO) NTN that feature large constellations with thousands of satellites to provide global broadband access [2]. The focus of this article is on airborne NTN utilizing the same frequency bands as ground based International Mobile Telecommunications (IMT) base stations (BS). This concept is known under the designation High Altitude Platform Stations (HAPS) as IMT base stations, or HIBS. By using the same spectrum as already identified for IMT and where deployments already exist today, HIBS can extend the operator's coverage area and benefit from the already existing device ecosystem.

While HAPS and HIBS both refer to High Altitude Platforms, they differ in the type of spectrum that they will be using. Often, the term HAPS is used also for the aircraft carrying the communications payload. Throughout this article, however, we reserve it (and the term HIBS) for the complete communications platform. For the aircraft alone, we use the term "high-altitude platform".

HIBS operate in the stratosphere, usually at an altitude of about 20 km. When compared to a terrestrial network, a HIBS system may provide wider coverage. When compared to a satellite network, a HIBS system may provide lower latency. Thus, in addition to satellite systems, HIBS can play a role for expanding mobile coverage to remote communities.

The studies of high-altitude platforms for telecommunications and remote sensing can be traced back to the 1990s [3]. A milestone for HAPS initiatives was the identification of the first frequency bands for HAPS in the Fixed Service in the International Telecommunication Union (ITU) Radio Regulations in 1997, followed by the identification of additional frequency bands for HAPS to provide IMT service (HIBS) in 2000. Despite the early interest, the development of high-altitude platforms for commercial connectivity in the 1990s and 2000s was limited due to immaturity of technical solutions. A resurgence of interest in providing connectivity using high-altitude platforms started around 2014, mainly driven by the Internet companies Google and Facebook that invested in new technology to beam connectivity through the atmosphere to reach remote areas. Admittedly, technology advancements in connectivity, as well as in areas such as solar panel efficiency, power storage, lightweight composite materials, avionics, microelectronics, and antennas have now made HAPS/HIBS systems more viable, leading to the creation of HAPS Alliance[†].

A high-level overview of HAPS communications was provided in [4] back in 2007. A more recent survey on high-altitude platforms was presented in [3], which focused on technologies directly related to airborne platforms but did not address the connectivity aspect. A comprehensive survey of the communication aspects can be found in [5]. The work [6] presented an investigation into the constellation design methodology of HAPS systems. A method for maximizing sum rate in a HAPS system, based on interference alignment, was proposed in [7].

---
[†] https://hapsalliance.org

|  | **Airplane** | **Balloon** | **Airship** |
|---|---|---|---|
| **Heavier or lighter than air** | Heavier than air | Lighter than air | Lighter than air |
| **Steerability** | Fully steerable | No or limited steerability | Fully steerable |
| **Power source** | Solar powered | | |
| **Operation altitude** | About 20 km | | |
| **Technical challenges** | Large wingspan needed, fragile construction | Limited steerability, cannot easily be flown to their area of operation | Large ground infrastructure, thermal management |

**Table 1** *High-altitude platform types*

Another study [8] used matching-game-based algorithms to find the optimal matching between users, HAPS, and LEO satellites. The integration of HAPS and satellites into a multi-layered network was also investigated in [9]. Finally, [10] suggested to use HAPS as communications platforms with integrated computing resources, while [11] highlighted HAPS as a potential enabler for next-generation parcel delivery networks.

In parallel with the academic studies about HAPS/HIBS, there have been notable new developments in standardization and on the regulatory front. In particular, the World Radiocommunication Conference 2019 (WRC-19) defined an agenda item for WRC-23 on HAPS as IMT BS, i.e., HIBS [12]. In addition, the 3rd Generation Partnership Project (3GPP) has been working on evolving the fifth generation (5G) radio access technology – known as New Radio (NR) – to support NTN which include HIBS systems [13].

Our article provides a concise but comprehensive survey of technical, regulatory, and standardization-related aspects of HIBS. Unlike earlier works, we focus on HIBS, because their relatively straightforward integration with terrestrial networks makes them a prime target for early commercial adaptation. We also provide some novel insights from state-of-the-art simulations of a combined HIBS and terrestrial network, to shed some light on the system level performance of such a combined network.

The remainder of this article is organized as follows. We start off by introducing the use cases of HIBS and the key characteristics of the different types of HIBS aircraft. We continue by discussing the spectrum aspects of HIBS and provide an overview of the HIBS standardization effort in 3GPP. Finally, we present system level performance evaluation results on HIBS, and conclude by highlighting some interesting directions for future research. The basic HIBS architecture is described in 3GPP TR 38.811: The HIBS payload serves devices on the ground via a service link, while being connected to the core network through a gateway via the so-called feeder link. While the feeder link is an integral part of a HIBS system, the focus in this article is on the service link. In particular, the HIBS frequency bands described below refer to the service link only.

## Use Cases of HIBS

This section outlines some of the potential use cases for HIBS.

*Network coverage expansion*: HIBS can cover sparsely populated or hard to reach geographical areas where terrestrial infrastructure is impossible or too costly to build (e.g. mountains, deserts, oceans, etc.). With the expected wide coverage from HIBS solutions, it might be possible to expand the network coverage area in a timely and cost-effective manner in order to narrow the infrastructure gap between urban and remote areas.

*Disaster resiliency*: Since HIBS provide connectivity from the stratosphere, they represent a highly resilient network infrastructure against natural disasters such as earthquakes, floods, and bushfires. A HIBS-based network can be used as a back-up network in the situation where terrestrial infrastructure is not operative as a consequence of a natural disaster. This can ensure a high-level of availability for critical operations.

*Fostering IoT deployment*: The deployment of mobile networks traditionally follows the population density. A HIBS-based network could provide coverage for IoT devices and goods, considering their more diverse uses and locations (crop, forest, wildlife, oil & gas, logistics). HIBS-based networks could complement terrestrial networks to expand the IoT deployment.

*Drones*: As HIBS are operating from the stratosphere, they can provide 3D coverage of a large geographical area to facilitate drone operations. Terrestrial networks may not be optimal for serving drones, due to their antennas being down-tilted to optimize coverage on the ground. A HIBS-based network can provide connectivity from the ground up to the sky.

## Characteristics of High-Altitude Platforms

Overall, high-altitude platforms can be classified into three groups with different characteristics and challenges: airplanes, balloons, and airships. Airplanes need to move through the surrounding mass of air to generate lift and thus need part of their energy to power their engines. Balloons and airships, on the other hand, are lighter than air and thus do not need to spend energy to stay airborne. Airships are equipped with engines and can be flown to their desired operation areas.

|  | ITU Region 1 | ITU Region 2 | ITU Region 3 |
|---|---|---|---|
| **HIBS frequency bands as per RR (Res.221)** | 1885-1980 MHz | | |
|  | 2010-2025 MHz |  | 2010-2025 MHz |
|  | 2110-2170 MHz | 2110-2160 MHz | 2110-2170 MHz |
| **Frequency bands for consideration under AI 1.4 at WRC-23 (Res.247)** | 694-960 MHz | | |
|  | 1710-1885 MHz | | |
|  | 2500-2690 MHz | | 2500-2655 MHz |

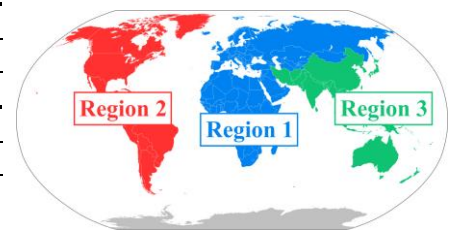

**Table 2** *HIBS spectrum overview*

Balloons, on the other hand, move only with the winds, resulting in limited steerability at best. Common to these three classes is that the vehicles are generally unmanned and fully automated, which is a prerequisite to allow economical operation and flight durations of weeks or even months.

The operational altitude of about 20 km is chosen mainly because of the favorable atmospheric conditions in this part of the atmosphere. Almost all weather phenomena happen below, and in particular wind speeds are very low and comparable to those at the surface. An additional benefit is that the airspace above approximately 20 km height is not regulated by air traffic control in most countries [3].

A further requirement for long-duration missions is that the vehicles must not depend on any carried fuel and thus are solar-powered. At the same time, this highlights one of the technological challenges of HIBS: operation at high latitudes during the winter months, with little sunlight, is challenging. The current generation of HIBS is restricted to a latitude band of approximately 35 degrees north and south of the equator, if year-round operation is desired.

Engineering challenges are plenty in high altitude platforms, and each vehicle type has its own. Since the lift force generated by an airplane's wing is proportional to the density of the surrounding air, which is reduced to about 5% at 20 km height, these airplanes need to employ extreme lightweight construction methods and an extremely large wingspan. The largest models have a wingspan of more than 70 m but weigh only a few hundred kg. The payload capacity is typically in the range of tens of kg. A benefit of the large wing is a large area that can be covered with solar panels.

So far, the airplane category of high-altitude platforms has been explored the most by the industry, and several different prototypes have been built and flown during the past 25 years. In 2018, the Airbus Zephyr was the first high-altitude platform to enter serial production, albeit still in single-digit numbers. An Airbus Zephyr plane also holds the record for the longest flight by any airplane with almost 26 days in 2018. The AeroVironment Sunglider (a descendant of the NASA Pathfinder/Helios program) also completed its first test flight in 2019. It is of particular interest because it is developed in a joint venture with the Japanese mobile network operator SoftBank and is intended to be used as a HIBS.

The world's first HIBS operator, however, was Loon, providing commercial 4G LTE service in Kenya from 2019 until their shut down in 2021. Loon did not use airplanes, but high-altitude superpressure balloons, which are a proven technology and allow extremely long flight durations. In 2019, one of Loon's balloons set the record with a flight duration of 223 days. At the end of the flight, they are landed and can be recovered. The biggest disadvantage of balloons is their relative inflexibility since they are not steerable and cannot easily be flown to a new region when demand changes. However, Loon developed methods to keep the balloons within their desired area of operation by utilizing different wind directions at different heights [14].

Airships, finally, seem at a first glance to combine the strengths of airplanes (their steerability) with those of balloons (their ability to stay airborne without consuming electrical power and their robustness). However, to allow for a reasonable payload mass, they need to be of very substantial size (typical lengths are >100 m), and thus require large and expensive ground infrastructure. Thermal modeling of such large airships is rather complicated and not yet fully understood. So far, only very few prototypes have been built, and these have been plagued by technical problems. The only project in active development seems to be the Stratobus by Thales Alenia Space.

The main characteristics of the three types of high-altitude platforms are summarized in Table 1.

Apart from the difficulties related to the construction and operation of the platforms, operating a communications payload in the stratosphere is a challenge in itself. In addition to the strict weight and power limitations, the equipment might need to cope with low temperatures (below -60°C), low air pressure, and high levels of UV radiation. This might put requirements on the electronics that are not very different from operating a satellite payload in space. Another challenge that is shared with communication satellites is the large area that has to be served by a single base station, limiting the possible user densities. The distance between base station and terminals is also large, which constrains the link budget, even though not as much as in the satellite case. One issue that is more severe as for satellites, however, is the fact that the elevation angle under which a terminal sees the HIBS will be lower due to the lower altitude compared to a satellite. While satellite systems often have a minimum elevation angle of around 45 degrees, HIBS might need to operate at elevation angles as low as 15 degrees if they should serve an area that is significantly larger than a typical terrestrial cell in a rural environment.

## Spectrum Aspects

Besides the technical challenges described in the previous section, HIBS need to follow spectrum regulations that are different from those of terrestrial networks. High-altitude platforms may provide user traffic in specific bands according to the ITU Radio Regulations (RR) [15]. Some of these bands are allocated to the Fixed Service (referred to as HAPS in the RR), while others are allocated to the Mobile Service (i.e. HIBS). It should be noted that this distinction is a regulatory one. As such, there is not necessarily a technological difference in HAPS vs. HIBS. However, Fixed Service is targeting ground stations at fixed locations, which enables the use of antennas with higher directivity than what is possible in mobile devices such as smartphones. This is still true with the introduction of ESIM (Earth Stations In Motion) in the Fixed Service, referring to ground stations on vehicles such as ships or airplanes.

With regards to HIBS, in particular the bands 1885-1980 MHz, 2010-2025 MHz and 2110-2170 MHz in Region 1 (Europe, the former Soviet Union, Africa, and the Middle East) and Region 3 (most of Asia and Oceania) and the bands 1885-1980 MHz and 2110-2160 MHz in Region 2 (North and South America) are identified in the RR as bands which may be used by high altitude platform stations as base stations to provide IMT service. Downlink transmission from HIBS is only allowed in the bands 2110-2170 MHz in Region 1 and Region 3 and 2110-2160 MHz in Region 2.

Resolution 221 of the RR stipulates the necessary technical conditions for HIBS for the purpose of protecting co-channel services and applications in adjacent countries, including terrestrial IMT. In addition, conditions to protect adjacent services are required. A notification of the frequency assignment of these stations to the Radiocommunication Bureau is also compulsory.

The WRC-19 defined an Agenda Item (AI 1.4) for the upcoming conference, WRC-23, on the use of HIBS in certain frequency bands below 2.7 GHz already identified for IMT, and, in particular, the bands 694-960 MHz, 1710-1885 MHz and 2500-2690 MHz. In preparation for WRC-23, the first meeting of the Conference, CPM23-1, tasked the ITU Working Party 5D to study the spectrum needs for HIBS as well as performing sharing and compatibility studies to ensure protection of other services in these bands. The results of these studies will be the basis for any decision to be taken at WRC-23 in relation to AI 1.4. Table 2 gives an overview of the existing and planned HIBS frequency bands.

ITU has identified a number of bands at higher frequencies for use by HAPS, e.g. at 6 GHz and 28/31 GHz. These are allocated to the Fixed Service, however, and thus not applicable for HIBS.

## HIBS Initiative in 3GPP

3GPP Release 15 contains the first complete specifications of 5G NR, paving the way for large-scale commercial 5G deployments. The Release 15 NR specifications are also the basis for the continuous evolution of 5G technology. Enabling 5G NR to support NTN is one direction under exploration in 3GPP.

In Release 15, 3GPP studied scenarios, requirements, and channel models for 5G NR based NTN, documented in 3GPP TR 38.811. In Release 16, 3GPP continued the effort by examining the solutions for adapting 5G NR to support NTN, summarized in 3GPP TR 38.821. Though the NTN work in 3GPP has been primarily focused on satellite-based radio access networks, HIBS systems have also been considered when applicable. Specifically, to evolve 5G NR to support NTN, there are four main challenges that shall be addressed: long propagation delays, large footprint sizes, moving cells, and pronounced Doppler effects. These challenges are all more significant in satellite-based radio access networks than in HIBS systems. Although there were no specific analyses conducted for HIBS systems during Release 16, it is expected that the same enhancements considered for satellite-based radio access networks may be applicable for HIBS systems when needed.

|  | **Terrestrial network** | **HIBS network** |
|---|---|---|
| **Channel model** | TR 38.901 (Rural Macro) | TR 38.811 (Rural) |
| **Inter-site distance** | 9 km | N/A |
| **Beam footprint size** | N/A | ~10 km |
| **Height** | 30 m | 20 km |
| **Carrier frequency** | 2 GHz | 2 GHz |
| **Carrier bandwidth** | 20 MHz | 20 MHz |
| **BS antenna model** | TR 36.873 (Three-sector sites) | TR 38.811 (Bessel function), with 16.5 dBi antenna gain |
| **BS transmit power** | 49 dBm | 49 dBm |
| **BS receiver noise figure** | 5 dB | 5 dB |
| **UE antenna gain** | 0 dBi ||
| **UE transmit power** | 23 dBm ||
| **UE receiver noise figure** | 9 dB ||
| **Data traffic model** | Full buffer ||

**Table 3** *Simulation assumptions*

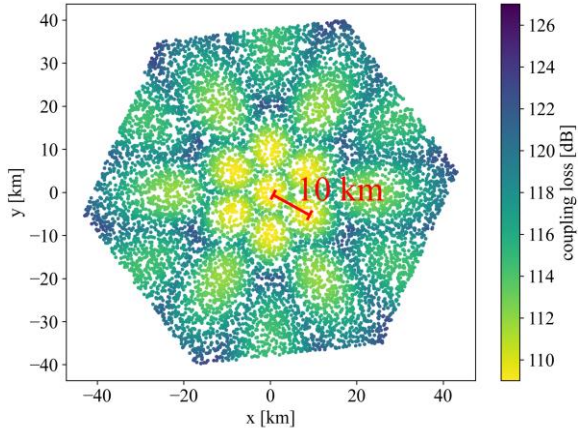

**Figure 1** *HIBS coupling loss*

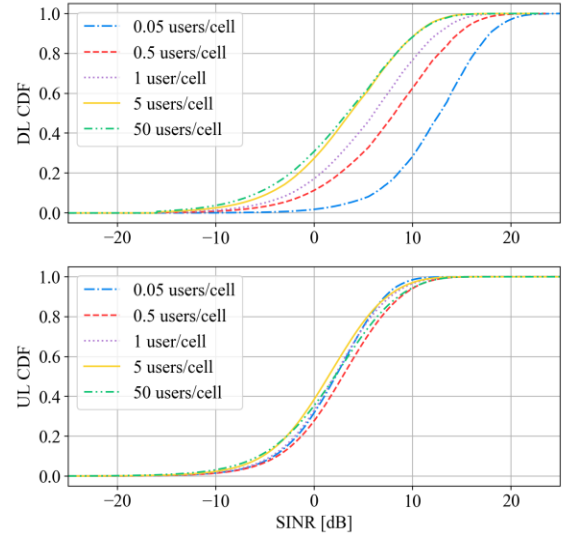

**Figure 2** *HIBS Downlink and Uplink SINR*

5G NR has been designed with terrestrial communications in mind. However, 5G NR is a flexible air interface, and it was found that existing functionalities form a good basis for supporting NTN deployments.

To address the unique challenges in NTN, 3GPP has conducted a work item to develop standardization enhancements to evolve 5G NR to support NTN in 3GPP Release 17 [13]. The work item aims to introduce enhancements for LEO and geostationary orbit (GEO) satellite-based radio access networks while being compatible for supporting HIBS systems and air-to-ground (A2G) communication.

## System Level Performance of HIBS

In this section, we present initial system simulation results to shed light on the performance of HIBS. Our simulations are based on the assumptions described in 3GPP TR 38.811, including the overall geometry of the setup, the antenna modeling, and the modeling of the radio propagation through the atmosphere. We simulate 19 cells, served by a single HIBS at an altitude of 20 km and an elevation angle of 90° as seen from the central cell. The beam footprint diameter of the innermost cell is approximately 10 km, resulting in a total service area of roughly 4000 km$^2$. Table 3 lists the most important simulation assumptions.

Figure 1 shows the coupling loss distribution across the beam footprint, including the 16.5 dBi antenna gain of the HIBS antenna. It can be observed that the coupling loss values in outer cells are much higher than those in central cells. This is a purely geometric effect and a result of the comparatively low altitude of a HIBS (vs. a satellite), which leads to a large variation in the HIBS-to-ground distance between the center and the outer regions of the coverage area. In the example simulated here, the distance to the central cell is 20 km (the flight altitude of the HIBS), while the distance to the outer edge of the simulated area is larger than 40 km. As a result, the outermost layer of cells shows a coupling loss that is at least 4 dB larger than that of the central cell. To counteract this effect, applying a different antenna beam pattern with a higher gain or a higher output power for the outer cells may be considered.

Figure 2 shows the signal-to-interference-and-noise (SINR) distribution for different user densities. We simulate full-buffer data traffic, i.e. all users are receiving (DL) or transmitting (UL) data all the time. In the DL, it can be observed that the SINR degrades with increasing user density. When the user density is below 1, there is on average less than one user per cell. In this regime, as the user density increases, inter-beam interference increases rapidly, leading to rapid degradation of the SINR. When the user density increases further, the system on average approaches a fully loaded state and enters an interference-limited scenario, making SINR insensitive to the user density.

The UL SINR does not depend on user density. Compared with the DL SINR, it is always low, even with very low user densities. The reason is that we assume a regular handheld and thus power-limited terminal. As a result, the UL SINR is noise-limited and not much affected by the increased interference resulted from increased user density. To reach better SINR, a terminal with higher transmit power or equipped with a high gain antenna could be considered.

We also study throughput performance for a HIBS network. In order to assess the service quality with respect to a terrestrial network, we extend the simulation setup with a terrestrial network (TN) of 12 three-sector sites arranged in a ring around a central coverage hole. The central area is served by a HIBS. For the sake of simplicity, we restrict the HIBS network to a single cell, roughly equivalent to the center cell of the previous simulation. The inter-site distance of the terrestrial network is 9 km, corresponding to a rural environment. This setup reflects the main use case of HIBS mentioned above: expanding network coverage to remote areas. Figure 3 shows the simulation setup and the user association to the terrestrial and HIBS cells.

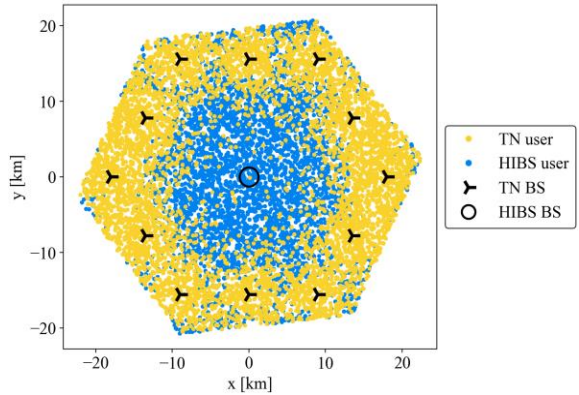

**Figure 3** *User association to a combined network of 12 terrestrial three-sector sites and a single HIBS cell*

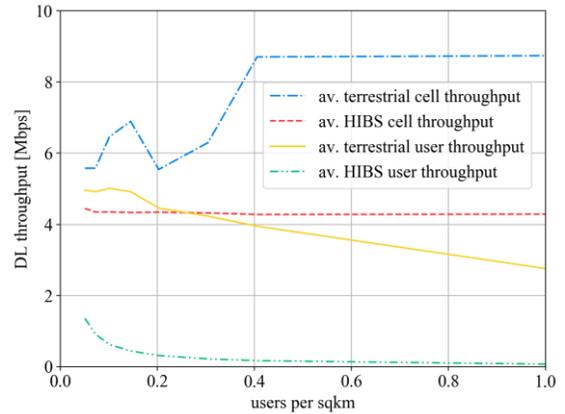

**Figure 4** *Downlink user and cell throughput*

The resulting user and cell throughput is shown in Figure 4 as a function of the user density. The HIBS cell throughput is relatively constant at about 4.2 Mbps, indicating that the cell is saturated already at very low user densities. The average terrestrial cell throughput fluctuates around 6 Mbps for very low user densities, before also saturating at close to 9 Mbps. The HIBS user throughput is very low. At the lowest user densities, it reaches above 1.5 Mbps, but falls quickly to around 100 kbps with increasing user density. The terrestrial user throughput also decreases when the available capacity has to be shared by more and more users, but it stays on a significantly higher level than in the much larger HIBS cell, which has to serve more than 10 times as many users, which are also located at much larger distances. The maximum spectral efficiency of the HIBS and terrestrial network is 0.23 bps/Hz and 0.45 bps/Hz, respectively.

Finally, we investigate user mobility. With the same network setup as before, we create users deep in the coverage of the terrestrial cells and let them move towards the center of the simulation area, where only the HIBS cell provides coverage. We also simulate the opposite case, where users are created within the HIBS cell and move towards the terrestrial coverage area. Figure 5 shows the locations where users have been handed over from the terrestrial to the HIBS cell (red) and vice versa (green). The shaded areas show the approximate coverage areas of HIBS (blue) and terrestrial network (yellow), respectively. In the TN-to-HIBS direction, the handover happens approximately at the intended border area. In the reverse case, however, users regularly move far into the area covered by the terrestrial network before the actual handover is performed. This reflects the differences in signal propagation, where the HIBS signal strength decays only very slowly with increasing distance from the cell center, and indicates that it might be beneficial to adapt the mobility procedures accordingly.

## Conclusions and Research Directions

Connectivity everywhere and at any time is critical and mobile networks are key to achieve this goal. 5G is not just an evolution of mobile networks but a revolutionary technology that will further advance society. As part of 5G, NR includes non-terrestrial connectivity. In particular, HIBS can expand the current mobile network operators' coverage area to remote areas where terrestrial deployments are challenging or impossible, while reusing mobile spectrum assets and taking advantage of the available ecosystem.

In this article, we have presented a state-of-the-art primer on HIBS, covering use cases, key characteristics of HIBS systems, spectrum aspects, 3GPP standardization, and system level performance evaluations. Our results illustrate how the coupling loss between a HIBS and a terminal on the ground varies across the beam footprint. The results also show that, in the DL, the SINR decreases with increasing user density. In the UL, in contrast, the SINR is always low because of the power limited terminals. Results from our mobility simulations show that users moving from a HIBS cell to a terrestrial network might be handed over later than necessary, due to the slowly decaying HIBS signal strength.

We conclude by pointing out some fruitful avenues for future research.

*Power consumption optimization*: Aerial platforms in HIBS systems usually rely on solar power systems to supply the necessary power for operations including telecommunications transceivers and antennas. As a result, the power availability might vary across daytime and nighttime and over months/seasons. Adapting and optimizing the communications design to the power constraint of HIBS systems is a largely under-explored research area.

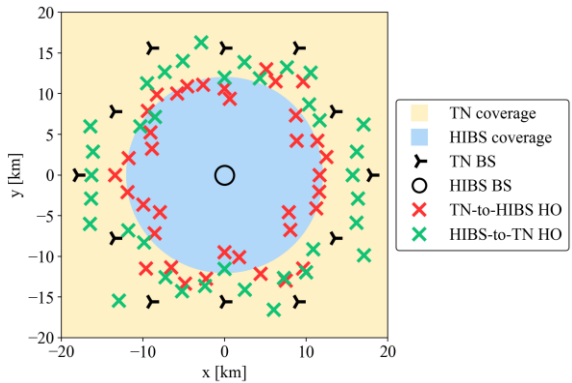

**Figure 5** *User mobility between terrestrial and HIBS cells*

*Connectivity for high-latitude regions*: Aerial platforms in HIBS systems relying on solar power may face challenges when flying at high-latitude regions during the winter months when daylight hours are few. How to provide connectivity to high-latitude regions is an important challenge to overcome.

*Coordination and coexistence between HIBS systems and terrestrial mobile networks*: As HIBS systems aim to use the same frequency bands as ground-based IMT BS, how to coordinate HIBS systems with terrestrial networks is an important and rich area to look into. Example issues include interference coordination, seamless mobility, and load balancing. Adjacent channel coexistence with neighboring mobile operators is also a key area for further investigation.

*Design aspects for feeder link:* Gateway transceivers are stationary, powerful, and typically dimensioned to make feeder links non-limiting. Consequently, the focus of this article is on service links. Nevertheless, there are design aspects for the feeder links that deserve investigation. For example, the spectrum used by the feeder link may overlap with the spectrum used by other services such as fixed satellite service. In such spectrum sharing scenarios, interference coordination between HIBS systems and other types of systems is an interesting topic. In some scenarios, the feeder of a HIBS system can come from a satellite link rather than the ground gateway, leading to a "space-sky-ground" integrated network. How to jointly design the feeder link and other links in such an integrated network is an open question.

*Trials and test deployments*: Due to the unique characteristics and challenges of deploying HIBS in the stratosphere, it is imperative to conduct extensive early trials and test deployments to collect feedback. Such trials and test deployments will help identify potential enhancement areas. The knowledge obtained can provide guidance to the evolution of 5G NR technology for better supporting HIBS systems.

Besides the technical and regulatory aspects described in this article, the realization of a commercial HIBS system faces many other challenges, e.g. to build the system in a cost-effective way, or to demonstrate the high reliability that is required. These aspects deserve further theoretical and practical analysis as well.